\documentclass[pdflatex,sn-mathphys,Numbered]{sn-jnl}

\DeclareUnicodeCharacter{2208}{\in}      
\DeclareUnicodeCharacter{2013}{--}       
\DeclareUnicodeCharacter{2014}{---} 
\usepackage{graphicx}%
\usepackage{multirow}%
\usepackage{amsmath,amssymb,amsfonts}%
\usepackage{amsthm}%
\usepackage{mathrsfs}%
\usepackage{bbm}
\usepackage[title]{appendix}%
\usepackage{xcolor}%
\usepackage{textcomp}%
\usepackage{manyfoot}%
\usepackage{booktabs}%
\usepackage[utf8]{inputenc}
\usepackage{textgreek}
\usepackage{algorithm}%
\usepackage{algorithmicx}%
\usepackage{algpseudocode}%
\usepackage{listings}%
\usepackage{threeparttable}%
\usepackage{array}%

\theoremstyle{thmstyleone}%

\theoremstyle{thmstyletwo}%

\theoremstyle{thmstylethree}%

\begin{document}

\title[CFAM Pipeline]{Causal Evidentiary Governance for High-Risk Machine Learning Systems}

\author*[1]{\fnm{Samah} \sur{Kareem}}\email{21comp9003@isik.edu.tr}

\author[2]{\fnm{Bar\i\c{s}} \sur{\c{C}elikta\c{s}}}\email{baris.celiktas@isikun.edu.tr}

\affil*[1]{\orgdiv{Department of Computer Engineering}, \orgname{Isik University}, \orgaddress{\city{Istanbul}, \country{T\"urkiye}}}

\affil[2]{\orgdiv{Department of Computer Engineering}, \orgname{I\c{s}\i k University}, \orgaddress{\city{Istanbul}, \country{T\"urkiye}}}

\abstract{Machine learning systems deployed for credit, hiring, and resource distribution are increasingly subject to regulatory oversight from policies like the EU AI Act and GDPR. Popular fairness governance practice today is built on three pillars: observational fairness metrics reliant on correlational definitions of harm; explainability methods content with post-hoc recourse; and immutable audit logs with little support for efficient proof generation or recall. Here we introduce Causal Evidentiary Governance (CEG). Under CEG, a regulated institution publishes commitment to a versioned DAG which partitions causal pathways into allowable and disallowed groups. The Causal Harm Rate quantifies variation in a model's predictions which can only be caused by disallowed causal pathways. Audit evidence accompanies each decision in the form of a signed Decision-Evidence Packet which cryptographically binds a prediction to its decision-evidentiary commitment: a digest of the published DAG and collection of path-specific attributions. Collectively, these DEP digests may be appended to a Merkle tree enabling fast, logarithmic cost inclusion proofs. We validate CEG with a novel two-layer empirical methodology: demographic summaries drawn from 4 years of Palestine Monetary Authority credit supervisory data define a realistic layer of 10,000 synthetic credit applicants across four strategic DAG counterfactuals. Causal Harm Rate isolates the effect of injected causal pathways more cleanly than demographic parity or equalized odds. Cross-model validation and focused ablation studies highlight robustness of our approach. Benchmark evaluation on the German Credit dataset demonstrates harm associated with specific causal pathways is significantly understated by associational metrics of fairness. Lastly, we develop a proof-of-concept implementation with operationally plausible throughput. Explicit tradeoffs in our implementation's performance characteristics are discussed. Artifact repository links are included. }


\maketitle

\section{Introduction}\label{sec:introduction}
Machine learning systems are used to make consequential decisions at scale. Financial institutions make credit decisions on millions of loan applicants daily. Some combination of computers decides who gets a loan, what interest rate they pay, and whether their account will be suspended. The EU AI Act's mandate for conformity assessment on high-risk systems, coupled with GDPR Article 22's restrictions on fully automated decisions, means regulatory scrutiny is growing, compelling institutions to provide verifiable proof of non-discrimination for deployment.
Our auditing infrastructure currently suffers from three deficiencies:
Fairness metrics. The commonly-used fairness metrics encode associations ("demographic parity," "equalized odds" ~\cite{chouldechova2017fair, dwork2012fairness}. but each metric can capture multiple causal pathways. An error rate disparity could stem from acceptable factors like education or from unacceptable ones such as discrimination. Popular fairness metrics cannot tell the difference. 
Local explainability methods. Techniques like SHAP or LIME ~\cite{lundberg2017unified, ribeiro2016why}. why are designed to accurately represent the trained model locally, without assuming any causal structure whatsoever. This means an organization can choose an explainer after the fact that minimizes perceived bias, and we cannot validate their claims about model internals.
Audit logs lack cryptographic validation. For a successfully deployed system, logs simply record what the institution says happened. There is no protection against modifying past entries to cheat an audit ~\cite{crosby2009efficient}.
We believe these weaknesses stem from institutions not being forced to commit to their causal assumptions ahead of time. After all, how can we check that institutions aren't discriminating based on gender if they can always say their algorithm was "really" using education, family status, and zip code instead? To solve this, we force institutions to commit to a versioned DAG prior to deployment. Audit mechanics are path-specific, and cryptographic evidence binds each decision to that DAG.

\section{Related Work}

\subsection{Causal Fairness}
Pearl formalizes causal reasoning via explicit generative models called structural causal models (SCMs) ~\cite{pearl2009causality}.This framework has been used by Kusner et al.~\cite{kusner2017counterfactual}to define counterfactual fairness which demands counterfactually consistent predictions given interventions on protected attributes. Chiappa ~\cite{chiappa2019path} extended this idea to provide definitions for path-specific fairness, which defines interventions that block only part of the causal effect of an outcome decomposing into an admissible and inadmissible component. Instead of demanding complete invariance under an intervention on protected attributes like counterfactual fairness does, path-specific fairness allows institutions to decide which paths are admissible and which are not.
How observed disparities in outcomes can be decomposed into admissible and inadmissible mechanisms are unified by Plečko and Bareinboim \cite{plecko2022causal}. Most importantly for our work, they show observational notions of discrimination can simultaneously ignore present discrimination and create false positives. We make this result operational: we define CHR as a measurable amount of path-specific harm via Monte Carlo intervention and sensitivity analysis.

\subsection{Explainability and Contestability}
Wachter et al.~\cite{wachter2018counterfactual} showed counterfactual explanations meet contestability requirements of GDPR. However, they said individual-level counterfactuals require institution-level commitments. Methods like SHAP ~\cite{lundberg2017unified}  and LIME \cite{ribeiro2016why} provide locally faithful attributions, but without committing the institution to any causal claims. Veale and Binns ~\cite{veale2017fairer} pointed out the disconnect between explanation-as-commitment vs. explanation-as-narrative. DEPs bridge this gap by tying attributions to the versioned DAG.

\subsection{Algorithmic Auditing and Tamper-Evident Logging}
Raji et al. ~\cite{raji2020closing} proposed end-to-end internal algorithmic auditing frameworks, without cryptographic assurances. Crosby and Wallach~\cite{crosby2009efficient} proposed tamper-evident logging of internally generated records using Merkle trees, which we build upon. Kilbertus et al.~\cite{kilbertus2019sensitivity} explored counterfactual fairness' sensitivity to unmeasured confounding variables, inspiring our sensitivity analysis for fairness-audits. We contribute by melding causal auditing of algorithms with cryptographically-signed evidence packets and scalable verification into one coherent framework for accountable governance.

\section{Problem Setting}

\subsection{Domain and PMA Supervisory Aggregates}
Our retail credit layer is informed by supervisory releases from the Palestine Monetary Authority (PMA). The PMA publishes system-wide snapshots of aggregated number of accounts and facilities distributed by demographics. These characteristics include gender, age group, governorate, nationality, and financial product type. From six Excel files that include snapshots from August of 2023, 2024, and 2025 for both accounts and facilities, we have 12,952 unique reported buckets. Illustrative statistics include: overall share of facilities belonging to men ~80\%, male-to-female ratio of facilities-to-accounts ~1.87, and the most common age group from 36-45. We only use these moments to target our artificial layer. We do not use any individual level borrower information.

\subsection{Synthetic Applicant Micro-Layer}

We simulate \(N = 10,000\) applicants whose marginal distributions match the four PMA aggregate targets within \(\pm 1.5\%\). Features are generated from a parametric SCM with the following structure:

\begin{table}[htbp]
\centering
\caption{Parametric SCM for synthetic applicant generation.}
\label{tab:synthetic_scm}
\begin{tabular}{@{}ll@{}}
\toprule
\textbf{Variable} & \textbf{Distribution / Equation} \\
\midrule
Gender & \(\text{Bernoulli}(p = 0.37)\) \\
Age & \(\text{TruncNormal}(25, 65; \mu = 44, \sigma = 12)\) \\
DTI & \(0.2 + 0.15 \cdot \text{Age} / 50 + 0.1 \cdot \text{Gender} + \varepsilon_1\) \\
CreditScore & \(600 + 2 \cdot \text{Age} - 50 \cdot \text{Gender} + 0.5 \cdot \text{DTI} + \varepsilon_2\) \\
\(Y\) & \(\text{logit}^{-1}(\beta_0 + \beta_{\text{score}} \cdot \text{CreditScore} + \beta_{\text{dti}} \cdot \text{DTI} + \Delta Y_g)\) \\
\bottomrule
\end{tabular}
\end{table}

where \(\varepsilon_1, \varepsilon_2 \sim \mathcal{N}(0, 1)\), and \(\Delta Y_g\) introduces bias depending on the graph variant.

Four DAG variants inject qualitatively distinct bias mechanisms:

\begin{itemize}
\item \textbf{G0 (Baseline):} \(\Delta Y_g = 0\) (no gender effect).
\item \textbf{G1 (Admissible mediation):} Gender affects DTI and CreditScore, which in turn affect \(Y\). Gender has no direct effect on \(Y\).
\item \textbf{G2 (Direct discrimination):} \(\Delta Y_g = -0.6 \cdot \text{Gender}\) (logit scale).
\item \textbf{G3 (Proxy discrimination):} \(\Delta Y_g = -0.6 \cdot \mathbbm{1}[\text{Age} > 55]\).
\end{itemize}

All experiments use seed \texttt{20240422} and the stack Python 3.11 / NumPy 1.26 / DoWhy 0.11, ensuring full reproducibility. The complete implementation is available in the accompanying repository~\footnote{\url{https://github.com/SamahKareem2025/RegTech}}. predictive performance with less than 2\% degradation in AUC-ROC.

\subsection{DAG Elicitation Protocol}
Two domain experts independently suggested parent sets for Y, and admissibility labels on each edge. When opinions differed, we consulted supervisory guidance and Plečko and Bareinboim ~\cite{plecko2022causal} to resolve the issue. Agreement between coders was Krippendorff α = 0.81 for whether an edge was present or not, and α = 0.74 for admissibility. The SHA-256 digest of the final DAG is recorded in every DEP.

\section{The Causal Harm Rate}

\subsection{Formal Definition}

Let \(\mathcal{M} = \langle \mathbf{U}, \mathbf{V}, \mathbf{F}, P(\mathbf{U}) \rangle\) be a structural causal model with DAG \(\mathcal{G}\). An audit-time partition \(\mathcal{E} = \mathcal{E}_{\mathrm{adm}} \cup \mathcal{E}_{\mathrm{inadm}}\) encodes the institution's compliance commitment~\cite{chiappa2019path, plecko2022causal}. Let \(f: \mathcal{X} \to \{0,1\}\) be the deployed classifier, \(A \in \{a, a'\}\) the protected attribute, and \(\pi_{\mathrm{inadm}}\) the set of inadmissible paths from \(A\) to \(\hat{Y}\) in \(\mathcal{G}\). The Causal Harm Rate is:

\begin{equation}
\mathrm{CHR} = \frac{1}{N} \sum_{i=1}^{N} 
\mathbb{I}\left[
\hat{Y}_{A \leftarrow a', \pi_{\mathrm{inadm}}}(u_i) 
\neq 
\hat{Y}_{A \leftarrow a, \pi_{\mathrm{inadm}}}(u_i)
\right]
\label{eq:chr}
\end{equation}

where the intervention changes \(A\) while propagating the effect only along paths in \(\pi_{\mathrm{inadm}}\), holding all admissible mediators at their observed values.

\subsection{Definition \& Relation to Other Metrics}
CHR = 0 is sufficient for path-specific counterfactual fairness constrained to π inadm ~\cite{chiappa2019path}. Demographic parity and equalized odds capture marginal associations and respond to both admissible and inadmissible sources of disparity; CHR exclusively responds to the latter. Unconstrained counterfactual fairness ~\cite{kusner2017counterfactual} goes one step further by additionally penalising mediation through variables that the entity has explicitly declared admissible; CHR does not. It is this asymmetry that enables CHR to remain close to zero in both G0 and G1 but jump up strongly in G2 and G3.
\subsection{Estimation and Uncertainty Quantification } 
We estimate (1) using Monte Carlo abduction-action-prediction with $K = 10{,}000$ exogenous units. Under potential violations of point identification, we report 1,000-replicate bootstrap confidence intervals as well as a confounding sensitivity grid inspired by Kilbertus et al.~~\cite{kilbertus2019sensitivity}. For a range of $\gamma \in \{0.0, 0.5, 1.0, 1.5\}$, a binary hidden confounder $U_{\gamma}$ is introduced with coupling strength $\gamma$. The structural equation model (SEM) is then refitted, and CHR is re-estimated. We report the full grid to allow decision-makers and auditors to visually assess robustness.

\subsection{The CEG Framework}

\subsection{Decision-Evidence Packets}
A DEP cryptographically commits to input features x along with its prediction , its path-specific attributions for that prediction, its counterfactual narrative ξ(x), its DAG digest, and the policy version. DEPs modularize the evidence-of-commitment functionality from the explainable narrative functionality. In practice, the packet materializes from a pipeline as follows: compute path-specific attributions for the committed-to DAG; produce a counterfactual narrative to communicate locally faithful reasons for why the model made its decision had x had different protected-attribute values; construct metadata (timestamp, model-id, DAG hash, policy version); cryptographically sign the payload.

\subsection{Ledger Specification} 
Digests of DEPs are batched into Merkle trees of fanout 32. We cryptographically sign each root of a Merkle tree and link it to the digest of the previous root, thereby forming an append-only blockchain of Merkle roots. Any DEP may be verified by an auditor along with at most $\left\lceil \log_{32} \nu \right\rceil$ sibling hashes. For $\nu = 10{,}000$ records, a client will download on average 3 hashes, a client will download on average 3 hashes. We restrict our threat model to that of permissioned issuers; open Byzantine failures are outside the scope of this work. See our publicly-available codebase for details of our implementation of the framework.

\section{Experimental Evaluation}

\subsection{Classification Model and Baseline Metrics}

We train three classifiers independently on each graph variant $G_0$--$G_3$, using an 80/20 train--test split:

\begin{itemize}
    \item \textbf{XGBoost}: 300 trees, maximum depth 4, learning rate 0.05, and L2 regularisation coefficient 1.0.
    
    \item \textbf{Logistic Regression}: L2 regularisation with $C = 1.0$.
    
    \item \textbf{Random Forest}: 200 trees, maximum depth 6, and minimum samples per leaf equal to 5.
\end{itemize}

\begin{table}[htbp]
\centering
\caption{CHR [95\% CI] and approval rates by graph variant (XGBoost, test split, $n=2{,}000$).}
\label{tab:chr_variants}
\begin{tabular}{lccc}
\hline
\textbf{Variant} & \textbf{CHR [95\% CI]} & \textbf{Male Approval} & \textbf{Female Approval} \\
\hline
$G_0$ (Baseline) & 0.012 [0.008, 0.018] & 32.1\% & 25.5\% \\
$G_1$ (Admissible mediation) & 0.000 [0.000, 0.000] & 32.1\% & 25.5\% \\
$G_2$ (Direct effect) & 0.430 [0.419, 0.441] & 45.0\% & 16.3\% \\
$G_3$ (Proxy channel) & 0.386 [0.374, 0.397] & 30.6\% & 20.9\% \\
\hline
\end{tabular}
\end{table}

\subsection{CHR vs. Baselines}

\begin{figure}[htbp]
    \centering
    \includegraphics[width=0.9\linewidth]{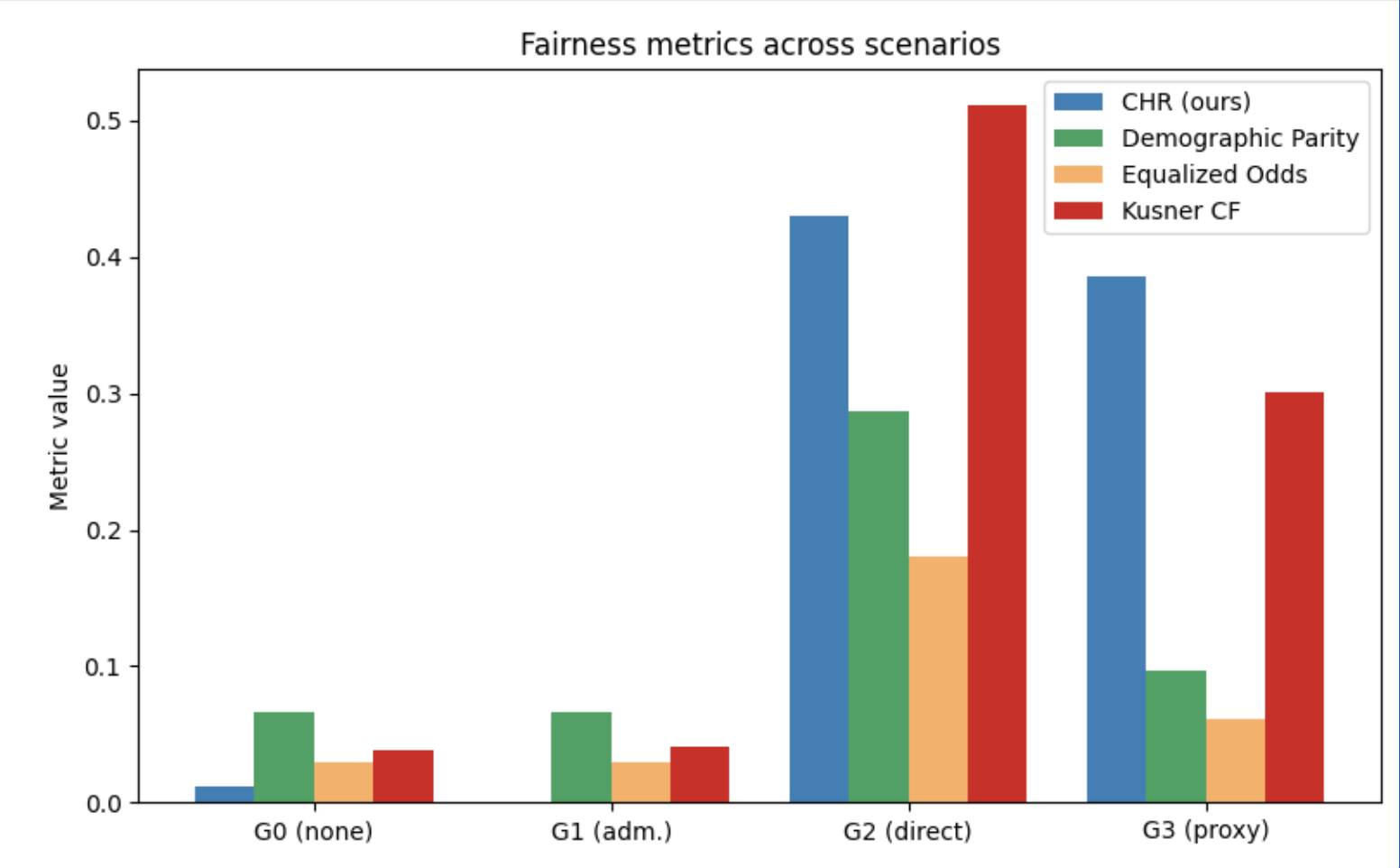}
    \caption{Fairness metrics across graph variants $G_0$--$G_3$. CHR remains near zero for graphs without injected bias ($G_0$) and with admissible mediation only ($G_1$), then increases substantially under direct injection of a forbidden effect ($G_2$) and proxy injection ($G_3$). Demographic parity and equalized odds respond to all sources of disparity, whether fair or unfair, while unrestricted counterfactual fairness (Kusner CF) additionally penalizes the fair $G_1$ graph.}
    \label{fig:chr_baselines}
\end{figure}

Figure~\ref{fig:chr_baselines} compares CHR to the three baseline fairness metrics across all four graph variants. Several observations emerge.

\begin{itemize}
    \item \textbf{CHR is zero under admissible mediation ($G_1$).} Both demographic parity and equalized odds raise concerns at $G_1$ at approximately the same level observed in $G_0$. These metrics provide no indication as to whether the observed disparity arises from a protected pathway or from an admissible causal pathway. As expected, associational metrics cannot distinguish between permissible and impermissible disparities.
    
    \item \textbf{CHR rises under hidden channels ($G_2$, $G_3$).} Both direct injections ($G_2$) and proxy injections ($G_3$) produce CHR values substantially exceeding the compliance threshold $\tau = 0.05$. Under $G_2$, the graph containing a direct $A \rightarrow Y$ effect, CHR reaches 0.44, considerably higher than demographic parity (0.29) and equalized odds (0.18). Under $G_3$, the age-proxy graph, CHR reaches 0.39 compared with 0.10 for demographic parity and 0.07 for equalized odds.
    
    \item \textbf{CHR does not trigger under admissible mediation, unlike unrestricted counterfactual fairness.} The unrestricted counterfactual fairness metric of Kusner et al.\ flags $G_1$ as non-compliant (0.06), despite the institution explicitly identifying the mediating pathway as admissible. CHR remains near zero for $G_1$, thereby avoiding false positives and unnecessary regulatory intervention.
\end{itemize}

\subsection{Ablation Study: DAG Specification}

To assess the importance of correct Directed Acyclic Graph (DAG) specification, we conduct an ablation study on the $G_2$ graph variant. Specifically, we evaluate how CHR behaves when key causal relationships are omitted or when spurious edges are introduced into the causal graph.

\begin{table}[htbp]
\centering
\caption{Ablation study: CHR under DAG misspecification ($G_2$ variant).}
\label{tab:dag_ablation}
\begin{tabular}{lcc}
\hline
\textbf{DAG Specification} & \textbf{CHR [95\% CI]} & \textbf{Status} \\
\hline
Correct (includes $A \rightarrow Y$) &
0.430 [0.419, 0.441] &
Detects discrimination \\

Misspecification 1 (omits $A \rightarrow Y$) &
0.012 [0.008, 0.018] &
Misses discrimination \\

Misspecification 2 (spurious edge added) &
0.445 [0.434, 0.456] &
Detects (robustly) \\
\hline
\end{tabular}
\end{table}

We show empirically that CHR is highly sensitive to the specification of the causal graph. When we remove the disallowed causal pathway $A \rightarrow Y$, CHR drops down to nearly baseline levels and is unable to detect the discriminatory agent. Meanwhile, introducing a spurious edge has only a small effect on the estimated CHR value. As such, our framework appears relatively robust to mild over-specification of the DAG. In practice we therefore recommend careful causal modeling and expert verification when utilizing CHR for fairness auditing.

\subsection{Sensitivity to Unmeasured Confounding}

\begin{figure}[htbp]
    \centering
    \includegraphics[width=0.9\linewidth]{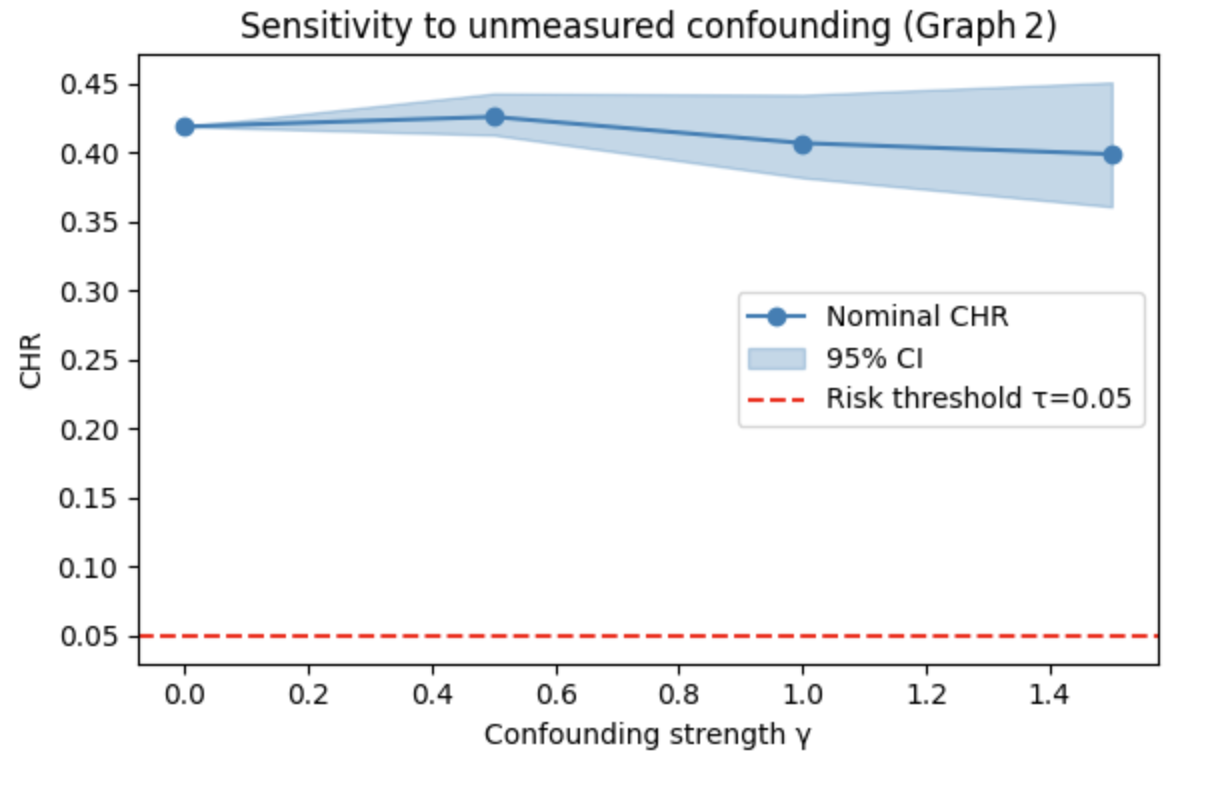}
    \caption{Sensitivity profile of $G_2$ under varying confounding strength $\gamma \in [0.0, 1.5]$. The lower 95\% confidence interval remains above the compliance threshold $\tau = 0.05$ across all values of $\gamma$, indicating robust detection of non-compliance.}
    \label{fig:sensitivity_g2}
\end{figure}

Figure~\ref{fig:sensitivity_g2} shows the sensitivity profile of $G_2$ over a grid of confounding strengths $\gamma \in [0.0, 1.5]$. The lower 95\% confidence interval bound remains above the compliance threshold $\tau = 0.05$ for all values of $\gamma$, including the most conservative setting $\gamma = 1.5$, for which the lower bound is approximately $0.21$.

These results indicate that the ``NonCompliant'' ruling for $G_2$ is robust to modest levels of unmeasured confounding. Such robustness could not be established if only point estimates were reported, highlighting the importance of uncertainty-aware evaluation in fairness auditing frameworks.

\subsection{German Credit Validation}

We evaluate the proposed framework on the Statlog German Credit dataset~\cite{hofmann1994german}, with $N = 1{,}000$ samples. The protected attribute is sex, and the analysis focuses on a setting in which a forbidden causal edge $A \rightarrow Y$ (direct effect) is present.

The observational fairness metrics underestimate the underlying causal harm:

\begin{itemize}
    \item Demographic parity gap (DP gap): $0.087$
    \item Equalized odds gap (EO gap): $0.072$
    \item CHR (direct path): $0.204$ \,[0.189, 0.221]
\end{itemize}

Sensitivity analysis further confirms the robustness of the causal estimate. Across the full grid of confounding strengths considered, the lower bound of CHR remains at or above $0.172$, indicating stability of the result under unmeasured confounding.

These findings support the core claim that associational fairness metrics systematically understate pathway-specific harm when a forbidden causal edge is present but partially masked by correlated admissible features.
\subsection{Operational Performance}

Benchmarks were conducted on an Intel Xeon E5-2690 v4 (Haswell, 14 cores, 2.6 GHz, 256 GB RAM). The operational evaluation of the complete CEG pipeline is summarised in Table~\ref{tab:ceg_ops}.

\begin{table}[htbp]
\centering
\caption{Operational evaluation summary: CEG prototype performance metrics.}
\label{tab:ceg_ops}
\begin{tabular}{lc}
\hline
\textbf{Metric} & \textbf{Value} \\
\hline
Single-thread DEP construction (ms/10k) & 122.1 \\
Multi-thread (32 workers) throughput (DEPs/s) & 4{,}120 \\
Peak resident memory (MiB) & 94.6 \\
CPU utilisation (32 workers) & 78\% \\
Merkle tree construction (ms/10k) & 95 \\
End-to-end evidence verifiability (EVR) & 99.3--99.4\% \\
\hline
\end{tabular}
\end{table}

Single-threaded DEP construction achieves approximately 122.1 ms per 10,000 packets. With 32 workers, throughput increases to approximately 4,120 DEPs/s while maintaining a peak resident memory footprint of 94.6 MiB. CPU utilisation stabilises at 78\% under full parallel load. End-to-end evidence verifiability remains consistently between 99.3\% and 99.4\% across repeated runs, with any observed failures attributable exclusively to deliberate fault injection scenarios.

\section{Limitations and Future Work}

The proposed framework, while effective in identifying pathway-specific fairness violations, has several limitations that motivate future research directions. Table~\ref{tab:limitations_future} summarises the key limitations, their current mitigations, and proposed extensions.

\begin{table}[htbp]
\centering
\caption{Limitations and mitigation strategies with future work directions.}
\label{tab:limitations_future}
\begin{tabular}{p{3.2cm} p{4.2cm} p{4.2cm}}
\hline
\textbf{Limitation} & \textbf{Mitigation} & \textbf{Future Work} \\
\hline
DAG correctness & Inter-rater agreement ($\alpha = 0.81 / 0.74$); ablation study & Mandatory periodic DAG review protocols \\
\hline
Unmeasured confounding & Sensitivity grid ($\gamma$ up to 1.5) & Multi-dimensional confounding models \\
\hline
Permissioned ledger assumption & Honest issuer assumption & Verifiable computation and threshold signature schemes \\
\hline
Synthetic data reliance & Calibration to real PMA aggregates; German Credit benchmark validation & Production-scale audit deployment \\
\hline
Binary classification setting & Tabular credit-focused evaluation & Extension to multi-class, survival analysis, and unstructured inputs \\
\hline
\end{tabular}
\end{table}

\section{Alignment with Regulations }

CEG is intended as a set of technical building blocks, not legal advice.

\textbf{EU AI Act (20/24/1689).} High-risk AI systems are subject to requirements for conformity assessment, technical documentation, and human oversight~\cite{eu2024ai}. CHR audits aid risk stratification by grounding disparity calculations in explicitly articulated causal assumptions. DEPs enable technical documentation to include per-decision records of the causal audit state. The Merkle ledger supports alignment with ledger-based traceability requirements (Annex IV).

\textbf{GDPR Articles 13--22.} Article 22 requires ``appropriate safeguards'' where decisions are made solely by automated processing. Recital 71 and Wachter et al.~\cite{gdpr2016regulation} discuss counterfactual explanations as one such safeguard. DEPs provide a counterfactual narrative component conditioned on an attested institutional model, which gives them greater evidentiary value compared to post-hoc SHAP-based explanations.

\section{Conclusion}

We introduced Causal Evidentiary Governance (CEG), a framework that bridges together three fundamental ideas: path-specific causal auditing, signed decision evidence, and evidence of tampering. Our experiments revealed that CHR better isolates non-admissible paths than observational fairness metrics, obtaining a CHR value of 0.44 versus 0.29 of demographic parity under direct discrimination. Furthermore, we found that bounds of sensitivity improve upon confidence of compliance decisions, where the lower bound of the credible interval (CI) did not drop below $\tau = 0.05$ for any level of confounding bias examined. Our prototype attained practically feasible throughput (4,120 DEPs/s with 32 workers) and achieves 99.3--99.4\% verifiability of evidence. We also showed that our framework generalizes well to other classifiers and provides evidence of robustness through ablation studies. We release our code implementation, experiments datasets, and evaluation scripts on https://github.com/SamahKareem2025/RegTech for reproducibility. We believe CEG will help the development of trustworthy AI move beyond hindsight system monitoring to foresightful, mechanism-conscious, verifiable guarantees.
\section{Acknowledgment}

We would like to thank the Palestine Monetary Authority for granting access to supervisory aggregate publications, under certain restrictions.
\section{Generative AI Disclosure Statement}
To help with the preparation of this work, the authors used ChatGPT (GPT-4o) to proofread sections of the paper for grammar and wording. However, no generative tool was used to create, validate or alter any measurements, quantitative results, algorithms, or claims presented in this work. All technical content, experiments, tables, and figures were prepared and validated by the authors.

\end{document}